\PassOptionsToPackage{hidelinks}{hyperref}
\documentclass[10pt,conference]{IEEEtran}

\IEEEoverridecommandlockouts

\usepackage{graphicx}
\usepackage{caption}
\usepackage{hyperref}
\usepackage{booktabs}

\usepackage{array}
\usepackage{calc}
\usepackage{seqsplit}
\usepackage{stfloats} 
\usepackage{placeins} 
\usepackage{float} 
\usepackage{url}
\usepackage{subcaption}
\usepackage{framed} 

\renewcommand{\baselinestretch}{0.929}  
\usepackage[compact]{titlesec}

\makeatletter
\renewcommand{\@biblabel}[1]{}
\makeatother

\title{Between the Commits: Process, Error, and Claim Reliability in a Wholly AI-Authored Codebase}

\author{
  \IEEEauthorblockN{D. J. Leith\thanks{Email: doug.leith@tcd.ie}}
  \IEEEauthorblockA{Trinity College Dublin\\Dublin, Ireland}
}

\newlength{\cslhangindent}
\newlength{\csllabelwidth}
\newenvironment{CSLReferences}[2]
 {\begin{list}{}{%
  \setlength{\itemindent}{0pt}
  \setlength{\leftmargin}{0pt}
  \setlength{\parsep}{0pt}
  \ifodd #1
   \setlength{\leftmargin}{\cslhangindent}
   \setlength{\itemindent}{-1\cslhangindent}
  \fi
  \ifnum #2 > 0
  \setlength{\itemsep}{#2\baselineskip}
  \fi
 }}
 {\end{list}}

\newcommand{\CSLLeftMargin}[1]{\parbox[t]{\csllabelwidth}{#1}}
\newcommand{\CSLRightInline}[1]{\parbox[t]{\linewidth - \csllabelwidth}{#1}\break}

\providecommand{\citeproctext}{}

\begin{document}

\maketitle

\begin{abstract}
We present: (i) a new dataset consisting of the full development history of a 21,000-line Python tool built entirely by Claude AI, with no human-authored code or tests, (ii) two code-provenance tracing tools, (iii) three taxonomies for instruction intent, commit provenance, and response reliability, (iv) application of these to analyse the dataset. We find that: (i) user coding agent CLI instructions differ in kind from IDE-chat instructions, with a greater focus on comprehension, planning and consultation, (ii) code development is mainly proactive, (iii) 14.3\% of AI code-generation events contain a real error later caught by the AI-authored test suite, (iv) roughly 1 in 4-5 of the AI's interactive responses contains one or more factual errors.
\end{abstract}

\section{Introduction}\label{introduction}

AI coding agents write code quickly, but make mistakes often enough that their output cannot be trusted at face value {[}1{]}, {[}2{]}. Judging how reliable AI-driven software development actually is requires visibility into the development process itself e.g.~how often the AI's own code and claims turn out to be wrong, and what catches those errors before they ship.

To gain that visibility we need full session transcripts and code development artefacts (e.g.~AI generated to-do lists) not just commit history. Unfortunately, existing large-scale datasets of AI-authored code consist primarily of commits/PRs with, at best, thin session coverage. Additionally, attributing a piece of code to its origin within an interactive session, recovering AI coding errors that were caught and fixed mid-session, and separating the AI's reliable claims from its unreliable ones each need tooling that does not yet exist.

The main contributions of this paper address these gaps: (i) the Extractor dataset; (ii) two reusable tools that trace AI-authored Python code back to its provenance; (iii) three taxonomies, each extending or filling a gap in an existing classification scheme, for instruction intent, commit provenance, and response reliability; and (iv) application of these to Extractor, to gain insight into the AI code development process.

The dataset contains full AI session transcripts, the complete commit history and AI-authored process artefacts (a running to-do list, a heuristics registry, its own limitations note) generated during the development of Extractor. Extractor is 21,000 lines of production code with a comparably sized test suite, built over 210 commits and 25 sessions from 678 user instructions. It was entirely built using Claude AI models over a period of several weeks. No code, test, or commit in its history has human authorship. The Extractor dataset provides a level of completeness no existing large-scale dataset offers (\S\ref{sec:related-work}).

The second contribution comprises two tools that trace AI-authored code back to its provenance. The first is a control-flow-graph block decomposition combined with \texttt{git\ log\ -L} history tracing, bridging renames. This traces individual code changes back to a specific user instruction, discovery, or self-correction event. The second is a session-transcript mining technique that recovers self-corrected code-generation errors from tool-use logs. These errors leave no trace in the commit history.

The third contribution is a set of three taxonomies. Each extends or fills a gap in an existing classification scheme, and each is of independent interest beyond the specific analysis it supports here. For the behavioural intent behind each user instruction, we extend Tang et al.'s IDE-chat taxonomy {[}3{]} to a multi-turn CLI setting. Their taxonomy explicitly excludes CLI-based coding agents. We add two categories that single-session chat has no need for: cross-session coordination, and correction of the AI's own conduct rather than its code. For commit provenance, we extend the standard Perfective/Corrective/Adaptive maintenance taxonomy {[}4{]}, {[}5{]} to separate why a change was made from how it was worked out. For the reliability of the AI's own claims to the user, we develop a taxonomy with no direct precedent: to the best of our knowledge, no prior work fact-checks an AI coding agent's multi-turn conversational claims against the software artefacts they describe.

We apply these tools and methods to the Extractor dataset. We find that user CLI instructions differ in kind from IDE-chat instructions, with a greater focus on comprehension, planning and consultation. Code development is mainly proactive rather than reactive, with 86-90\% of commits and 89-90\% of code blocks tracing to a proactive trigger rather than being a reaction to a bug or test failure. We find that 14.3\% of Claude AI-generated code contains errors that are caught by the AI-authored test suite before any commit (information that is unobservable by standard commit-based analysis). In interactive coding agent sessions we find that the Claude AI's responses are relatively reliable when reporting or verifying a fact (94.3\% accurate), less reliable when explaining an existing mechanism (91.1\%) or diagnosing a root cause (89.3\%), and substantially worse when proposing a design fix (79.4\%). Overall, we find that roughly 1 in 4-5 of AI interactive responses contain at least one inaccuracy. As far as we are aware, these are the first reported error rates for user-facing responses for a modern coding agent.

\section{Related work}\label{sec:related-work}

\begin{figure}[tb]
\centering
\includegraphics[width=1\linewidth,height=\textheight,keepaspectratio]{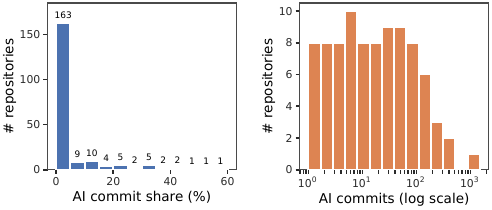}
\caption{Lifetime AI-authored commit share (a) and absolute count (b) per repository across all 205 repositories in the SWE-chat dataset (on 21st Aug 2026); the 117 repositories with zero AI-authored commits are omitted from (b)'s log-scale axis.}\label{fig:swechat-ai-commit-distributions}
\end{figure}

\textbf{Datasets.} The closest existing dataset is probably SWE-chat {[}6{]}, linking session transcripts to commit-level human/AI attribution across 205 real repositories, but session coverage per repository is thin (median 8 sessions, 17.6\% of repositories have only one logged session) and lifetime AI-authored commit share is negligible for most (median 0\%, maximum 55.6\%, 57.1\% have no AI-authored commits at all; see Figure \ref{fig:swechat-ai-commit-distributions}). AIDev {[}7{]} is a larger PR/commit-level dataset (932,791 agent-authored pull requests across 116,211 repositories, five agents including Claude Code), but captures only PR and commit metadata, not session transcripts, and its cutoff (1 August 2025) predates recent improvements in AI code-writing\footnote{In November 2025 Claude Opus 4.5 was the first model to break 80\% on SWE-bench Verified.}.

\textbf{Tools.} \emph{Provenance attribution.} SZZ, introduced by {[}8{]}, traces a bug back to its introducing commit via \texttt{git\ blame} on the fixed lines. It is known to misattribute tangled or reformatting-only commits as the true origin. Recent refinements add outside evidence to resolve this using bug-discussion text to disambiguate the relevant file {[}9{]}, or an LLM to judge the correct candidate commit {[}10{]} . This paper's own CFG-block decomposition plus \texttt{git\ log\ -L} tracing (\S\ref{sec:cfg-tracing}) targets a different, broader problem at finer granularity: every code block's full origin and churn history, not one bug-fix's introducing commit. Its own residual gap is different in kind: a rename-detection blind spot on partial-file moves, not commit misattribution, closed mechanically rather than with an LLM, to keep the tool's output reproducible.

\emph{Trajectory mining.} In {[}11{]} and {[}12{]}, the single step that causes a benchmark-task trajectory to fail overall is localised (on Terminal-Bench and SWE-bench-family tasks respectively). In {[}13{]} process differences between resolved and failed runs are profiled using fully deterministic anti-pattern detectors, deliberately excluding an LLM judge from its main claims. All three treat failure as the run never reaching a working solution at all, not a single test failing partway through and later being fixed, the gap our session-transcript miner is built to close.

\textbf{Taxonomies.} \emph{Instruction intent.} In {[}14{]} 11 developer-AI interaction types (auto-complete, command-driven, conversational, event-triggered, etc.) are proposed, but this is a purely conceptual paper with no underlying usage data. Recently, {[}3{]} instead derive an empirical, 7-category/20-subcategory intent taxonomy for IDE-integrated chat. Their dataset excludes CLI-based agents, ``primarily Claude Code,'' explicitly leaving that interaction paradigm, this paper's own subject, to future work.

\emph{Commit provenance.} Existing commit-taxonomy work classifies why a change was made after the fact, from the message or an issue thread {[}4{]}, {[}15{]}, {[}16{]}, and rationale-mining work does the same from Jira discussion or the message itself {[}17{]}, {[}18{]}. We are not aware of prior work that classifies a commit's trigger from the agent's own reasoning and tool use as it happened, rather than reconstructing it afterwards.

\emph{Response reliability.} Prior fact-checking work mostly focusses on journalism, social-media and LLM long-form text fact-checking {[}19{]} and verification is via internet search. Our own decompose-classify-verify pipeline follows a similar high-level architecture to ClaimBuster's four-stage pipeline {[}20{]} and SAFE's fact-decomposition-and-verification approach {[}21{]}, but checking factual claims about software is rare in the literature and as far as we know, there is no prior work on fact-checking AI coding-session multi-turn conversational transcripts.

\textbf{AI analysis.} \emph{Code-generation error rate.} Prior measurements of how often AI-generated code is wrong come from two sources, neither of which can see an error caught and fixed before it is ever committed. Benchmark studies report pass/fail rates against curated problems, not real development e.g. {[}1{]}, {[}2{]}. Repository-level studies instead compare finished, already-committed code: {[}22{]} classifies defects and vulnerabilities in shipped AI-authored code using rule-based pattern matching, primarily on the older, ChatGPT-3.5-era HMCorp dataset; {[}23{]} and {[}24{]} compare committed AI- and human-authored diffs by code metrics and CodeQL-detected vulnerabilities, and by technical debt, respectively; {[}25{]} and {[}26{]} analyse GitHub review and maintenance patterns from PR metadata.

\emph{Code provenance.} The same repository-level studies also compare AI-authored code against a human-authored baseline. Extractor has none, the whole codebase is AI-authored, so this paper instead traces each code block back to the commit and trigger that produced it (\S\ref{sec:commits}), asking how the code came to be written, not how it compares to a human reference.

\section{Extractor}\label{extractor}

Extractor is a Claude Sonnet AI-written tool (no human-written code), written from scratch over several weeks of work, see Table \ref{tab:extractor-scale}. It is a real tool (for reverse engineering protobuf descriptors from compiled Apple Objective-C/C binaries), not synthetic test or ``toy'' code.

\begin{table}[tb]
\footnotesize
\centering
\caption{Extractor summary.}\label{tab:extractor-scale}
\begin{tabular}{@{}
  >{\raggedleft\arraybackslash}p{(\linewidth - 8\tabcolsep) * \real{0.2541}}
  >{\raggedleft\arraybackslash}p{(\linewidth - 8\tabcolsep) * \real{0.1303}}
  >{\raggedleft\arraybackslash}p{(\linewidth - 8\tabcolsep) * \real{0.1642}}
  >{\raggedleft\arraybackslash}p{(\linewidth - 8\tabcolsep) * \real{0.1847}}
  >{\raggedleft\arraybackslash}p{(\linewidth - 8\tabcolsep) * \real{0.2667}}
@{}}
\toprule
\begin{minipage}[b]{\linewidth}\raggedleft
Size (src/test LOC)
\end{minipage} & \begin{minipage}[b]{\linewidth}\raggedleft
Tests
\end{minipage} & \begin{minipage}[b]{\linewidth}\raggedleft
Commits
\end{minipage} & \begin{minipage}[b]{\linewidth}\raggedleft
Sessions
\end{minipage} & \begin{minipage}[b]{\linewidth}\raggedleft
User instructions
\end{minipage} \\
\midrule
21,000 / 21,829 & 1,467 & 210 & 25 & 678 \\
\bottomrule
\end{tabular}
\normalsize
\end{table}

\textbf{Purpose.}
Many of Apple's system daemons and frameworks on iOS communicate, and
persist data on disk, using Protocol Buffers {[}27{]}. Almost none of the
\texttt{.proto} schema files that describe those messages are shipped anywhere,
public or private, on the device. To read what a given binary actually
sends, receives, or stores, that schema has to be recovered directly from
the compiled code itself. Extractor (\texttt{apple\_protobuf\_extractor}) is a tool that automates that recovery: given an
Objective-C Mach-O binary or dyld shared cache, extract a complete,
correct \texttt{.proto} descriptor for every protobuf message class it defines.

\textbf{Main components.}
Two things make this challenging.
First, almost all of the protobuf traffic on the platform is serialised
by \texttt{PBCodable}, a private Apple Objective-C protobuf implementation. Behaviour has to be inferred from how the compiled code actually behaves.
Second, Objective-C's message-dispatch machinery
and Apple's stripped, private serialisation internals leave a decompiler
with almost nothing useful to render, so the schema has to be recovered
from raw ARM64 disassembly instead. That is slow and unpleasant
work when done by hand: it means tracing individual instructions, one
call site and one basic block at a time, across potentially thousands of
message classes in a single framework. It is this that motivated the development of the Extractor tool.

\textbf{Dataset.}
The Extractor dataset is available at https://github.com/doug-leith/extractor-code-development. This contains commits, session history transcripts, AI-generated note files, the taxonomy codebooks (instruction, commit, and response) and scoring rubric, raw LLM categorisation and verdict results, and scripts to regenerate the tables and figures here.

\section{Analysis of code development process}\label{sec:mechanisms}

\subsection{Types of instruction given}\label{sec:instructions}

In this section we characterise how the user directed and collaborated with the AI over Extractor's development. We classify each of the 678 user instructions\footnote{After data cleaning to exclude: 274 background task notifications/wakeups injected as user turns, 19 /compact commands.} given to Claude into behavioural-intent categories, following the taxonomy developed by Tang et al. {[}3{]} from single-session IDE chat transcripts. Their scheme captures what kind of thing an instruction is asking for, e.g.~new code, a bug report, a request to review output, independently of the specific project. We use it to characterise the division of labour between user and AI, and to test how far a taxonomy built from short IDE chat exchanges carries over to a CLI-based coding agent used across long, multi-session engagements.

\subsubsection{Decomposing instructions}\label{decomposing-instructions}

\begin{figure}[t]
\centering
\begin{minipage}{0.44\linewidth}
\centering
\scriptsize
\setlength{\tabcolsep}{2pt}
\begin{tabular}{@{}p{0.08\linewidth}p{0.68\linewidth}p{0.17\linewidth}@{}}
\toprule
\# & Component & Label \\
\midrule
1 & when tests pass, (i) commit changes & 5.2;6.3 \\
2 & (ii) update to-do list to mark \#54 and \#58 done and remove them from the active list. & 6.1 \\
3 & (iii) consider a separate question: \texttt{\seqsplit{proto/extractor.py}} is 1502 lines long, that seems pretty big. do you recommend splitting it or not? & 3.1 \\
4 & (iv) also check for other large py file. & 3.2 \\
\bottomrule
\end{tabular}
\caption*{(a) Instruction idx289}
\end{minipage}
\hfill
\begin{minipage}{0.52\linewidth}
\centering
\includegraphics[width=\linewidth]{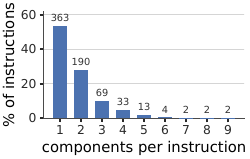}
\caption*{(b) Components per instruction}
\end{minipage}
\caption{Instruction decomposition into components. (a) An instruction decomposed into its components, each independently classified against the taxonomy codebook. (b) Components per instruction, nearly half (46\%) bundle two or more.}
\label{fig:decomp-examples}
\label{fig:components-per-instruction}
\end{figure}

In the Extractor corpus user instructions can be compound instructions, involving multiple steps. This is illustrated, for example, in Figure \ref{fig:decomp-examples}. As Figure \ref{fig:decomp-examples} shows, the steps can also involve instructions with different intents e.g.~runtime monitoring (``when tests pass''), toolchain use (``commit changes''), documentation (``update to-do list''). Similarly to Wei et al. {[}21{]}, we therefore use an LLM to decompose instructions into their component parts before categorising their intent\footnote{We found that decomposing instructions into components significantly improves classification consistency. For example, we drew a random sample of 50 instructions that decompose into more than one component, yielding 50 full instructions and 128 components. Using the final codebook, see \S\ref{sec:codebook-refinement}, we used Codex and Claude to classify the instructions in this sample with/without decomposition, allowing multiple categories to be assigned to each instruction or component instruction. The \(\kappa\) was 0.700 with full instructions vs 0.790 for decomposed instructions, with exact match rates of 52\% and 76.6\% respectively.}. Figure \ref{fig:components-per-instruction} shows that almost half of instructions are compound instructions, they are not rare.

\subsubsection{Codebook development}\label{sec:codebook}

Tang et al use an iterative abductive coding method to develop their taxonomy codebook. In this approach a sample of instructions is classified against the current codebook using an LLM, a human reviewer inspects each assignment, and where the fit is poor a new category is proposed. Tang et al.~converged over four rounds (9, 5, 1, 0 new or revised subcategories per round). We follow a similar abductive coding approach, modified as follows. We draw the first sample of instructions using stratified sampling rather than uniformly at random\footnote{To stratify, we roughly clustered the instructions using an LLM (``what themes do you see in the user instructions'') and sampled two instructions from each of the resulting eight clusters. The aim of the stratification is to allow us to quickly sample a wide diversity of instructions, so rough clustering is sufficient.}. Following Tang et al we categorise these instructions using an LLM instructed using: (i) Tang et al's codebook plus a temporary Others subcategory for instructions that did not fit an existing definition, and (ii) the instruction together with its immediate context (the preceding instruction and the AI's response). A human reviewer inspects each assignment, where the fit is poor a new category is proposed using corpus evidence and the codebook's definitions and boundary notes, and this is repeated until no new categories are proposed.

In the first round, three additional categories were judged necessary, in the second round one more category was added, in the third round no new categories were proposed. This abductive process yielded four additional categories: one under Context Specification (4.3 Design Principle Specification), one under Delegation (6.3 Version Control), and two under Workflow Control (7.6 Compliance Correction, 7.7 Session/Environment Coordination).

Tang et al explicitly exclude CLI agents like Claude Code from their data, and two of the new categories capture differences between CLI agents and the single-session IDE chats studied by Tang et al.~Claude Code sessions can run in parallel on the same project, unlike single-session IDE chats, and as a result there is a new category of user instructions handling cross-session co-ordination (7.7 Session/Environment Coordination). A second new category 7.6 Compliance Correction captures corrections to what Claude itself just did or displayed, not of the code under development e.g.~a failed or ignored instruction. The remaining two additions are not tied to how Claude Code works. 4.3 Design Principle Specification captures standing, codebase-wide quality statements; Tang et al.'s closest category is about interaction style, not code properties, so this appears to fill a gap in their scheme. Tang et al's 6.2 Toolchain Operation encompasses git commits etc, but we split these out into 6.3 Version Control because git instructions are a frequently occurring category in their own right. See Table \ref{tab:codex-taxonomy-distribution} for a summary of the resulting taxonomy.

\begin{table*}[tb]
\footnotesize
\centering
\caption{Instruction categories identified in Extractor corpus (full definitions in the codebook, dataset package).}\label{tab:codex-taxonomy-distribution}
\begin{tabular}{@{}
  >{\raggedright\arraybackslash}p{(\linewidth - 2\tabcolsep) * \real{0.2000}}
  >{\raggedright\arraybackslash}p{(\linewidth - 2\tabcolsep) * \real{0.8000}}
@{}}
\toprule
\begin{minipage}[b]{\linewidth}\raggedright
Category
\end{minipage} & \begin{minipage}[b]{\linewidth}\raggedright
Examples
\end{minipage} \\
\midrule
1.1 New Implementation & ``look into adding a check for use of google protobufs e.g.~for GPBMessage, so we can give the tool user an informative message''; ``then look into improving the test coverage so we can properly test later changes for regressions'' \\
1.2 Iterative Modification & ``then swap into production''; ``update tool to reflect this change'' \\
1.3 Alignment Correction & ``the reference protos are not ground truth, they're known to contain errors'' \\
2.2 Symptom Description & ``fix heuristic\_freq.py''; ``with locationd the regex gate fails with garbage.'' \\
3.1 Planning \& Consultation & ``plan out what step 1 would involve''; ``suggest develop a proof of concept first that can be tested'' \\
3.2 Project Comprehension & ``no field types with swift? is that unavoidable or a tool limitation?''; ``we already talked at length about enum-name-prefix and decided the enum type name was cosmetic, look back to see if you can find that conversation and confirm'' \\
3.3 General Knowledge Query & ``is that open source?''; ``while we wait, does the c++ libprotobuf-lite allow recovery of field names, numbers and types?'' \\
4.1 Information Injection & ``previously you advised waiting until after readFrom was updated, but I've decide to defer the readFrom update'' \\
4.2 Behavior Specification & ``without modifying production code''; ``ignore the test lines of code'' \\
4.3 Design Principle Specification & ``it should not depend on hardcoded memory addresses or the like - it needs to be portable across different binaries.''; ``it should have good test coverage.'' \\
5.1 Code Review & ``check that code has appropriate comments linking to each of the entries in the limitations doc''; ``then sample some of the repeated enums that changed and analyse the binary to confirm the changes are correct'' \\
5.2 Runtime Inspection & ``do this more complete test for the iphone8 data too.''; ``when its all ok proceed to do step 3 to complete the migration'' \\
6.1 Documentation & ``if not then update it''; ``add a comment to the code re the architectural constraint'' \\
6.2 Toolchain Operation & ``go ahead with (1)''; ``move the macho\_extractor apple folder to the project root'' \\
6.3 Version Control & ``then commit to git''; ``commit to git'' \\
7.1 Confirmation & ``ok, understood re proto rules''; ``ok, got it re the enum values.'' \\
7.2 Continuation & ``keep going''; ``continue'' \\
7.5 Sentiment Expression & ``ok i think i misunderstood.''; ``i like the behaviour of this session: you're careful and dogged'' \\
7.6 Compliance Correction & ``you've dropped the links to to-dos, add those back in.''; ``the column "row" should still be called "function" surely'' \\
7.7 Session/Environment Coordination & ``that sessions commits are now done.''; ``save in a file so i can pick up in another session'' \\
\bottomrule
\end{tabular}
\normalsize
\end{table*}

\subsubsection{Codebook refinement}\label{sec:codebook-refinement}

We refined the codebook to clarify boundary cases using an iterative process. We began by drawing 100 random
component instructions and classifying them using two independent LLMs (Codex GPT-5.6 Sol and Claude Sonnet 5, from two competing vendors) instructed using the behaviour intent codebook. As Tang et al.~note, and their own results support, there is good evidence that LLMs can achieve strong performance on text classification tasks when guided by structured instruction and schema constraints {[}3{]}. Gilardi et al.~report ChatGPT outperforming crowd-worker annotators accuracy by around 25\% on comparable text-annotation tasks {[}28{]}, and Zheng et al.~find a strong LLM judge can match human-human agreement levels on pairwise preference judgements, while also documenting potential biases (position, verbosity, self-enhancement) {[}29{]}. Disagreements in classifications between two independent LLMs therefore provides some insight into ambiguities in the boundaries between categories. For example, in ``Go ahead with Phase A'' the code for Phase A doesn't exist yet, which sounds like ``new'' work. But if the design was already fully worked out in an earlier conversation (a validated prototype, a detailed plan) and this instruction just says ``now actually build it,'' is that creating something new (1.1 New Implementation), or finishing something already decided (1.2 Iterative Modification). A human researcher reviewed disagreements between the two LLMs, looking for patterns in the boundary cases. For commonly occurring patterns the codebook was updated with a clarifying reminder. This was repeated for samples of 200, 300 and 600 components with convergence achieved after three iterations.

After this process, Cohen's \(\kappa=0.808\) between the LLMs (this \(\kappa\) is for the 512 instructions remaining after excluding the 166 instructions viewed during the codebook development and the iterative refinement process i.e.~it is for a held-out sample). This indicates that the LLMs are in good agreement and that the codebook establishes consistent boundaries between categories. It also shows that the results are not sensitive to the choice of LLM used to perform the classification.

To validate classifier accuracy, we drew a stratified sample of 55 instruction components, 5 instruction components from each category containing at least 25 instructions. A human researcher independently annotated these 55 instructions using the finalised codebook, yielding \(\kappa=0.710\) with Claude and \(\kappa=0.718\) with Codex. That is, the human categorisation is consistent with both LLM categorisations (\(\kappa\) in range 0.61--0.80 indicates ``substantial agreement'', and is somewhat higher than Tang et al's human-human \(\kappa=0.669\)).

The full codebook, including definitions, key signals, examples, and boundary notes, is provided in our dataset package.

\subsubsection{Results}\label{results}

\begin{figure}[t]
\centering
\includegraphics[width=0.26\linewidth,angle=90]{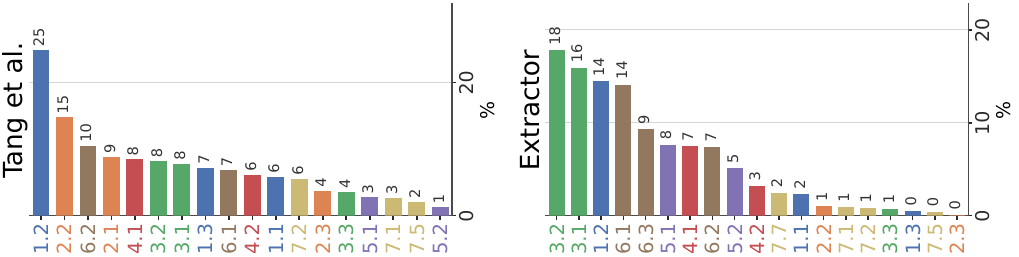}
\caption{Behavioral-intent subcategory distribution: Tang et al.'s published data (left, IDE-chat messages) vs. this study's Extractor corpus (right). Bars are coloured by top-level category, categories below 1\% in both panels are omitted.}
\label{fig:codex-taxonomy-distribution}
\end{figure}

Our taxonomy extends Tang et al.'s scheme, developed and validated on single-session IDE chat transcripts, to a CLI-based coding agent. Three new subcategories capture observed CLI-specific behaviour (cross-session coordination, standing design-principle statements, and correction of the AI's own session conduct rather than its code), and boundary-note refinement sharpened category distinctions the original codebook left underspecified in the CLI context.

Figure \ref{fig:codex-taxonomy-distribution} summarises the distribution of instruction categories observed in the Extractor corpus and also shows the corresponding distribution for Tang et al's corpus. It can be seen that the distributions are not the same. In the Extractor cli-based coding agent corpus Project Comprehension, Planning \& Consultation, Documentation and Runtime Inspection are much more prominent than in Tang et al's chat IDE corpus. Symptom Description is much less frequent in the Extractor corpus and Log Paste is absent. Iterative Modification, Information Injection and Toolchain Operation are similarly prominent in both corpora.

These differences mean that not all of Tang et al findings carry over unchanged to the Extractor corpus.

Finding 1 ``Conversational programming operates through progressive refinement rather than upfront specification'' is similar in that a substantial fraction of instructions relate to Code Authoring but that category 1.2 Iterative Modification is much more prevalent than category 1.1 New Implementation (6.0\(\times\) more here, vs.~4.2\(\times\) more in Tang et al.). However, category 1.3 Alignment Correction is much rarer in the Extractor corpus than in Tang et al.'s corpus, perhaps reflecting recent improvements in coding agent instruction following.

Finding 2 ``Developers delegated error diagnosis to AI by reporting symptoms and machine outputs rather than articulating code-level causes'' is not supported by the Extractor corpus. Failure Reporting, ``the second most common category'' in Tang et al.'s data, appears in only 1.77\% of instructions here (vs.~24.0\%), and Log Paste never occurs at all. Instead, in the Extractor corpus Validation is considerably higher (18.73\% vs 3.99\%), reflecting detection of errors by code review and runtime testing rather than by human inspection and input of failures. This appears to reflect a genuine difference in usage between IDE chat sessions and CLI-based coding agents.

Finding 3 that developers ``queried AI about behavior, outputs, and feature logic rather than reading code directly,'' is more true for Extractor than in Tang et al.'s data (Project Comprehension is 25.7\% vs.~8.2\%). Finding 4 ``Developers used AI to evaluate code through both static review and dynamic runtime inspection'' and Finding 5 ``documentation to externalize plans, explanations, and progress into persistent artifacts'' show a similar pattern i.e.~the qualitative finding holds, but at two-to-five times Tang et al.'s rate (Documentation 2.6\(\times\), Validation 4.7\(\times\)). Finding 6 that developers ``actively managed AI autonomy by shaping its context and constraining or delegating its behavior'' is similar with Information Injection (11.7\% vs.~8.5\%) and Behavior Specification (5.2\% vs.~6.1\%) both close to Tang et al.'s rates.

\subsection{Triggers for code development: Commit analysis}\label{sec:commits}

This section classifies each commit in the Extractor corpus by its
development trigger, e.g.~whether it was prompted by a test failure or made proactively, using a
taxonomy built from the AI coding agent's own session transcript, not just the commit message.
Existing commit-taxonomy work classifies why a change was made after
the fact, from the message or an issue thread {[}4{]}, {[}15{]}, {[}16{]}, and rationale-mining work does the same
from Jira discussion or the message itself {[}17{]}, {[}18{]}. We are not aware of prior work that classifies a
commit's trigger from the agent's own reasoning and tool use as it
happened, rather than reconstructing it afterwards.

We enriched raw code commits in two ways: (i) by finding the command making the commit in the session history, the session context (user instructions, Claude responses and actions) that led to the commit becomes available, (ii) often commit messages cite project documents, e.g.~\texttt{TODO.md}, \texttt{heuristics\_registry.py}, and these also provide additional context for the commit.

Using this enriched context, we used a similar approach to \S\ref{sec:codebook}--\S\ref{sec:codebook-refinement} to develop a taxonomy for the commits. We began by considering whether to decompose commits. Our analysis showed that 14 of the 210 commits merged multiple different changes. In view of this relatively small number, and the known difficulty of untangling merged commits {[}30{]}, we chose not to decompose the commits. A commit taxonomy was then developed using an iterative abductive approach: select categories iteratively with convergence after three rounds, then refine boundary cases by examining the differences between the categorisations by two independent LLMs (Codex GPT-5.6 Sol and Claude Sonnet 5). For the final codebook Cohen's \(\kappa=0.786\) between Claude and Codex.

Table \ref{tab:commit-type-definitions} summarises the final taxonomy. It is primarily structured around proactive and reactive code development. These high-level categories correspond to the Perfective and Corrective categories of Swanson (1976) {[}4{]}, carried forward into the current maintenance-process standard {[}5{]}. Swanson also defines an Adaptive category capturing code changes in response to changes in the software's environment e.g.~new OS, new platform, new regulatory requirement, but this has no analogue in the Extractor corpus. Conversely, our Housekeeping category, which captures common maintenance activity with no code-behaviour content, has no Swanson analogue. The proactive and reactive categories are sub-divided by method, i.e.~changes derived from (i) a specification/document, (ii) analysis of artifacts lacking documentation, (iii) a design decision, (iv) a code audit/review.\\
In the Extractor corpus reactive code changes are made in response to one of four triggers: cross-decoder differential (two independent decoding paths disagree on the same input e.g.~the tool's own decoder against a reference implementation, or two of the tool's own internal decode routes), systematic sweep (a broad scan across many binaries surfaces a pattern of anomalies not tied to any single known bug), external validator rejection (an external tool, e.g.~the protobuf compiler, rejects the tool's output as invalid), or unit/regression-test trigger (a test-suite run fails). Refactoring and addition of tests fall under Proactive-Design/Proactive-Audit.

The full codebook, including signals, examples, and boundary notes, is provided in our dataset package.

\begin{table}[tb]
\footnotesize
\centering
\caption{Definitions of the commit-type categories.}\label{tab:commit-type-definitions}
\begin{tabular}{@{}
  >{\raggedright\arraybackslash}p{(\linewidth - 2\tabcolsep) * \real{0.2200}}
  >{\raggedright\arraybackslash}p{(\linewidth - 2\tabcolsep) * \real{0.7800}}
@{}}
\toprule
\begin{minipage}[b]{\linewidth}\raggedright
Category
\end{minipage} & \begin{minipage}[b]{\linewidth}\raggedright
Definition
\end{minipage} \\
\midrule
Proactive-Spec & Content directly translated from a documented format spec/ABI (no investigation needed to write it). \\
Proactive-Analysis & Content found by directly inspecting a raw artifact, because no documented spec covers it (e.g.~disassembling compiled code). \\
Proactive-Design & Content is a general software-design/architecture decision. \\
Proactive-Audit & Found by systematically re-reviewing already-working code or tests for quality/robustness, via one of two sub-routes: general review (a self-review pass identifies a list of follow-up items) or necessity verification (a candidate mechanism is disabled/relaxed and the output re-diffed; zero diff proves it dead/redundant before retirement). \\
Reactive-Spec & Found reactively, the fix was determined from a documented format spec/ABI. \\
Reactive-Analysis & Found reactively, the fix required directly inspecting the raw artifact under study to determine. \\
Reactive-Design & Found reactively, the fix was a straightforward engineering change (no spec citation or artifact investigation needed to determine it). \\
Housekeeping & Docs, notes reorganisation, renames/archiving, build-metadata changes, a mechanical correction of a prior commit's own incomplete git operation, a commit that only records an empirical finding without changing any code status/behaviour. No code-behaviour content. \\
\bottomrule
\end{tabular}
\normalsize
\end{table}

\begin{table}[tb]
\footnotesize
\centering
\caption{Breakdown of the commit population by category and, separately, of CFG code blocks by the category of their originating commit.}\label{tab:commit-type}
\begin{tabular}{@{}
  >{\raggedright\arraybackslash}p{(\linewidth - 8\tabcolsep) * \real{0.3101}}
  >{\raggedleft\arraybackslash}p{(\linewidth - 8\tabcolsep) * \real{0.1600}}
  >{\raggedleft\arraybackslash}p{(\linewidth - 8\tabcolsep) * \real{0.1652}}
  >{\raggedleft\arraybackslash}p{(\linewidth - 8\tabcolsep) * \real{0.1800}}
  >{\raggedleft\arraybackslash}p{(\linewidth - 8\tabcolsep) * \real{0.1847}}
@{}}
\toprule
\begin{minipage}[b]{\linewidth}\raggedright
Category
\end{minipage} & \begin{minipage}[b]{\linewidth}\raggedleft
Commits (Codex)
\end{minipage} & \begin{minipage}[b]{\linewidth}\raggedleft
Commits (Claude)
\end{minipage} & \begin{minipage}[b]{\linewidth}\raggedleft
Code blocks (Codex)
\end{minipage} & \begin{minipage}[b]{\linewidth}\raggedleft
Code blocks (Claude)
\end{minipage} \\
\midrule
Proactive-Audit & 71 (33.8\%) & 76 (36.2\%) & 437 (23.6\%) & 673 (36.4\%) \\
Proactive-Design & 35 (16.7\%) & 31 (14.8\%) & 513 (27.8\%) & 476 (25.8\%) \\
Proactive-Analysis & 27 (12.9\%) & 31 (14.8\%) & 706 (38.2\%) & 496 (26.8\%) \\
Proactive-Spec & 0 (0.0\%) & 0 (0.0\%) & 6 (0.3\%) & 6 (0.3\%) \\
Housekeeping & 56 (26.7\%) & 56 (26.7\%) & -- & -- \\
Reactive-Analysis & 10 (4.8\%) & 7 (3.3\%) & 111 (6.0\%) & 117 (6.3\%) \\
Reactive-Design & 10 (4.8\%) & 8 (3.8\%) & 52 (2.8\%) & 57 (3.1\%) \\
Reactive-Spec & 1 (0.5\%) & 1 (0.5\%) & 23 (1.2\%) & 23 (1.2\%) \\
\textbf{Total} & \textbf{210} & \textbf{210} & \textbf{1848} & \textbf{1848} \\
\bottomrule
\end{tabular}
\normalsize
\end{table}

Table \ref{tab:commit-type} summarises the breakdown of commits by category by two independent LLMs. It can be seen that two LLMs are in good agreements. They both categorise the bulk of the commits as proactive, with a majority of those arising from audits and a roughly equal split between analysis and design for the rest. The reactive commit share (10.4--13.6\% across the two raters) is low by comparison with human baselines: below Lientz, Swanson and Tompkins' (1978) survey estimate of 17\% {[}31{]}, and toward the better end of the measured distribution across 7,557 GitHub projects (median 20\%, ranging from 6\% to 39\% across deciles) {[}32{]}, {[}33{]}.

\subsection{Triggers for code development: Code block analysis}\label{sec:cfg-blocks}

\begin{figure}[tb]
\centering
\includegraphics[width=0.95\linewidth,height=\textheight,keepaspectratio]{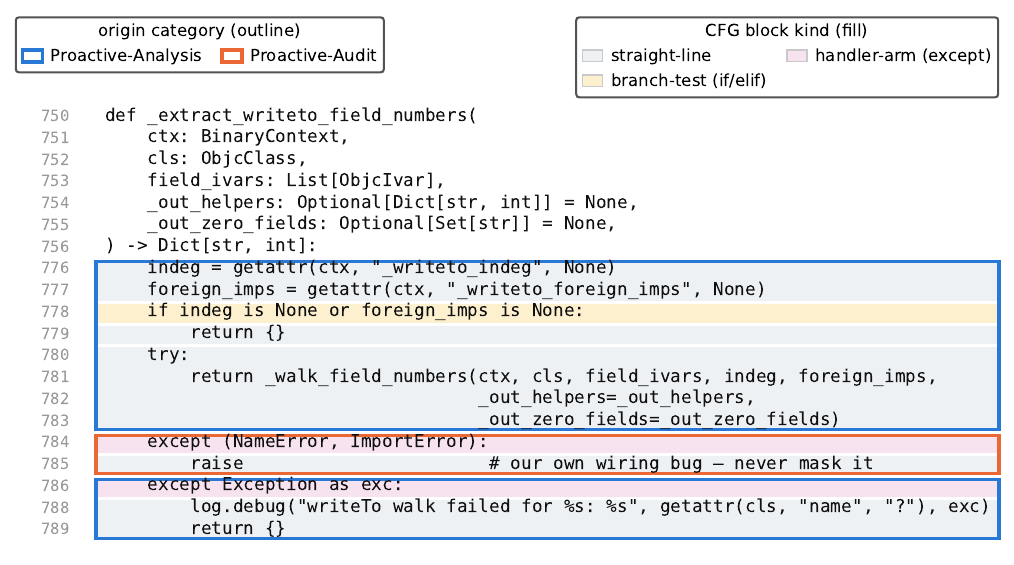}
\caption{Illustrating CFG code blocks and their provenance.}\label{fig:cfg-block-origin-example}
\end{figure}

Since the commits vary greatly in size, the commit-level breakdown in Table \ref{tab:commit-type} is rather coarse. We therefore also analyse code development using the codebase itself as the reference, giving the Code blocks/\% columns of the same table.

\subsubsection{CFG code block decomposition and tracing}\label{sec:cfg-tracing}

We built a tool to attribute every code block in Extractor to its originating commit and track its subsequent churn, without manual tracing or AI judgement. It has two parts: \texttt{cfg\_builder.py}, which partitions each function's abstract syntax tree (AST) into control-flow-graph (CFG) basic blocks, and \texttt{git\_tracer.py}, which recovers each block's commit history via \texttt{git\ log\ -L}.

This is a related but different problem from SZZ {[}8{]}: SZZ links one known bug-fixing commit back to the single commit that introduced the defect, at file/diff granularity via \texttt{git\ blame}; we instead attribute every code block, buggy or not, to its full origin and churn history at CFG-basic-block granularity, which no existing SZZ implementation does.

\begin{figure}[tb]
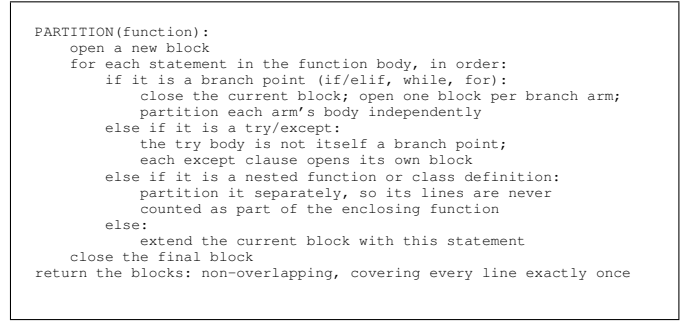

\fontsize{5.5}{6.5}\selectfont
\begin{framed}
\begin{verbatim}
PARTITION(function):
    open a new block
    for each statement in the function body, in order:
        if it is a branch point (if/elif, while, for):
            close the current block; open one block per branch arm;
            partition each arm's body independently
        else if it is a try/except:
            the try body is not itself a branch point;
            each except clause opens its own block
        else if it is a nested function or class definition:
            partition it separately, so its lines are never
            counted as part of the enclosing function
        else:
            extend the current block with this statement
    close the final block
return the blocks: non-overlapping, covering every line exactly once
\end{verbatim}
\end{framed}
\caption{Pseudo-code for CFG basic-block partitioning of a function.}\label{fig:cfg-partition-pseudocode}
\end{figure}

\textbf{Block partitioning.} Figure \ref{fig:cfg-partition-pseudocode} sketches \texttt{cfg\_builder.py}'s partitioning algorithm: it opens a new block at every branch point and every \texttt{except} handler, treats a nested function or class definition as a self-contained sub-problem so its lines are never attributed to the enclosing function, and otherwise extends the current block statement by statement.

\textbf{Provenance tracing.} Figure \ref{fig:cfg-tracer-pseudocode} sketches \texttt{git\_tracer.py}'s algorithm for recovering a block's origin commit and churn history. The straightforward case asks \texttt{git\ log\ -L} directly for the block's line-range history; the fallback case, needed when that returns nothing, is discussed next.

\begin{figure}[tb]
\fontsize{5.5}{6.5}\selectfont
\begin{framed}
\begin{verbatim}
TRACE_ORIGIN(block):
    ask git for the line-range history of the block's current location
    if git returns a history:
        origin = its earliest commit; churn = every later commit
    else:
        # the block may have arrived via a partial-file move that
        # git's own rename detection did not follow
        if the block's file was carved out of a larger file earlier:
            search that earlier commit's tree for the block's
            own text, to relocate it before the move
            if the text is not unique:
                instead search for the enclosing function by name
                (a stronger anchor than a short line of code), then
                confirm the block's text really sits inside that
                function's body, and that the commit which appears
                to have carved it out actually touches both the
                old and new file, before trusting the match
            if relocated:
                repeat the whole trace from the earlier commit
    if still unresolved: report it as unresolved, not a guess
\end{verbatim}
\end{framed}
\caption{Pseudo-code for tracing a code block's origin commit and churn history.}\label{fig:cfg-tracer-pseudocode}
\end{figure}
\textbf{The technical challenge.} \texttt{git\ log\ -L}'s rename detection is based on whole-file similarity, so it misses a block's history whenever only part of a large file was carved out into a new one, the dominant refactoring pattern in Extractor's history. We bridge this in two stages: first by relocating the block via its own text in the pre-carve file, and, when that text is too short or repeated to be unique (e.g.~a bare \texttt{return\ (x,\ x)}), by relocating it via its enclosing function's name instead. That second anchor is not always unique either: the search is a bare-name match against every function definition anywhere in the codebase at that point in history, so it fails, deliberately, whenever two different classes or functions happen to share a method name (\texttt{writeTo} and \texttt{readFrom}, for instance, are defined once per message class, so many classes share the same name). Rather than guess between candidates in that case, the tracer reports the block as unresolved. Even when the name is unique, a matching name and matching text alone are still not proof of common ancestry since an unrelated function elsewhere could coincidentally share both, so we additionally require the specific commit that appears to have carved the block out to genuinely touch both the old and new file before trusting the match. Across the full \texttt{proto/}+\texttt{macho/} sweep, adding the function-anchor stage cut the unresolved rate from 11.4\% to 0.4\%; every remaining unresolved block is disclosed with the reason it could not be relocated.

An LLM could plausibly relocate some of the remaining 1.9\% that our mechanical bridge leaves unresolved. We used the mechanical approach instead: the residual miss rate is already small, so there is little left to gain.

\subsubsection{Results}\label{results-1}

\begin{figure}[t]
\centering
\begin{minipage}{0.48\linewidth}
\centering
\includegraphics[width=\linewidth]{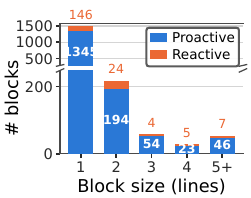}
\end{minipage}
\hfill
\begin{minipage}{0.48\linewidth}
\centering
\includegraphics[width=\linewidth]{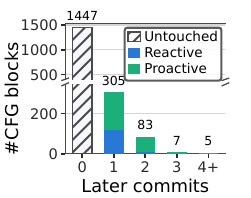}
\end{minipage}
\caption{CFG code blocks broken down (a) by size and proactive/reactive provenance, and (b) by number of later commits touching them and why.}
\label{fig:cfg-block-size-by-category}
\label{fig:churn-composition-histogram}
\end{figure}

Table \ref{tab:commit-type}'s Code blocks columns show the number of code blocks associated with each category of originating commit, separately for each rater. The large majority of code blocks originate proactively for both raters: 1662 (89.9\%) for Codex and 1651 (89.3\%) for Claude, with the remainder (186, 10.1\%, and 197, 10.7\%) reactive. Figure \ref{fig:cfg-block-size-by-category} shows the Codex proactive/reactive split broken down by code block size (the Claude breakdown is almost identical). It can be seen that the proactive/reactive split is similar regardless of block size (and that larger code blocks are relatively rare).

Proactive-Design, Proactive-Analysis and Proactive-Audit together account for the large majority of code blocks (Codex: 1656 of 1848, 89.6\%, Claude: 1645 of 1848, 89.0\%): proactive architecture/integration decisions and proactive investigation of the compiled artifact are both far more common, block-for-block, than any reactively-triggered category. Reactive-Analysis (a reactive trigger whose fix required directly inspecting the compiled artifact) is the largest reactive category.

\subsection{Code churn}\label{sec:code-churn}

Since we can trace code blocks back through historical commits, we can analyse code churn. Figure \ref{fig:churn-composition-histogram} summarises the frequency of changes made to each code block, broken down by their reactive/proactive provenance (housekeeping does not affect code). It can be seen that the great majority of code blocks undergo no changes. For code blocks that undergo one or two changes, just over half (270 of 400) are proactive. Reactive changes tend to be concentrated as one-off fixes. Code blocks with more than two changes are rare

\begin{figure}[tb]
\centering
\includegraphics[width=1\linewidth,height=\textheight,keepaspectratio]{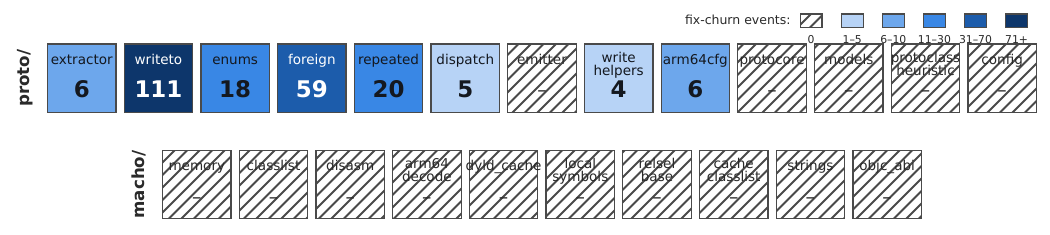}
\caption{Churn by Extractor module; hatched cells are modules with no churn.}\label{fig:bug-hotspot-map}
\end{figure}

Figure \ref{fig:bug-hotspot-map} maps code churn onto the modules within Extractor. This shows that most of the churn (79.2\%) is concentrated in \texttt{writeto.py}, \texttt{foreign.py} and \texttt{repeated.py}. These three modules do the primary reverse engineering work, and the churn here is consistent with the exploratory nature of that work already noted.

Note that while this churn analysis gives insight into the code development process, it does not provide much insight into Claude's coding accuracy. This is because obvious coding errors would be caught by tests etc before being committed and so would not show up in the commit history. Instead, we use session history data to analyse Claude's coding accuracy below.

\section{Reliability of Claude code generation}\label{sec:code-accuracy}

As already noted in Section \ref{sec:code-churn}, commit-based analysis misses code defects that Claude introduces but finds and fixes before the code is committed. We can, nevertheless, gain some insight into this by analysing the Claude session transcripts. These histories include details of tool use, including runs of pytest and the resulting output.

\subsection{Session-transcript mining}\label{session-transcript-mining}

Git history only records what was eventually committed; a bug Claude introduced, caught by a failing test, and fixed before ever committing leaves no trace there. Recent trajectory-mining studies of coding agents don't address this gap: they study benchmark-task trajectories that end in outright task failure, the agent never solving the assigned issue at all {[}11{]}, {[}12{]}, not a project's history of individual test failures caught and fixed along the way. To measure it, we mine the raw Claude Code session transcripts directly. Unlike the commit log, these preserve every tool call: the exact \texttt{pytest} command run and its pass/fail output.

For each session's raw \texttt{.jsonl}, our custom Python tool finds every Bash call containing \texttt{pytest}, classifies it PASS/FAIL by regex against its \texttt{-q} summary line, and flags a \texttt{FAIL\ -\textgreater{}\ (\textgreater{}=1\ edit)\ -\textgreater{}\ later\ PASS\ on\ an\ overlapping\ test\ target} sequence as a candidate self-correction episode. A companion pass tags every pytest invocation PASS/FAIL to also give a ``ran clean'' denominator. We found that raw session logs frequently contain duplicates since Claude re-emits earlier events each time a session is compacted, but these can be de-duplicated by using the event UUID. This part of the tool is purely mechanical Python and says nothing about why a \texttt{pytest} run failed.

Inspection of the session transcripts reveals deliberate mutation testing i.e.~mutate the code on purpose, confirm the test now fails, revert. This produces a \texttt{pytest} pattern identical to a genuine bug, i.e.~FAIL, then an edit, then PASS. Telling the two apart requires reading the session narration; a mechanical regex pass cannot distinguish them, and neither can trajectory-diagnosis tools built to stay fully rule-based for auditability {[}13{]}. We therefore instructed an AI (Claude Sonnet 5) to read each episode's transcript and assign it to one of four categories: (i) a production-logic defect, (ii) a test/fixture defect, (iii) a failure to propagate a change to a dependent test, or (iv) a deliberate mutation-testing cycle. Every episode categorised as a defect was then checked by a human, which caught two mistakes.

\subsection{Results}\label{results-2}

We analysed this data to identify cases where a failing run on a given test target is followed, later in the same session, by a passing test run on an overlapping target i.e.~candidate instances of code being written, found wrong, and corrected before commit, yielding the results shown in Table \ref{tab:selfcorrect}.

\begin{table}[tb]
\footnotesize
\centering
\caption{Definitions of the categories used in pytest analysis.}\label{tab:selfcorrect-definitions}
\begin{tabular}{@{}
  >{\raggedright\arraybackslash}p{(\linewidth - 2\tabcolsep) * \real{0.3496}}
  >{\raggedright\arraybackslash}p{(\linewidth - 2\tabcolsep) * \real{0.6504}}
@{}}
\toprule
\begin{minipage}[b]{\linewidth}\raggedright
Category
\end{minipage} & \begin{minipage}[b]{\linewidth}\raggedright
Definition
\end{minipage} \\
\midrule
Clean (no failure) & A \texttt{pytest} invocation reporting zero failures. The code under test worked as checked, whether on the first attempt or a later confirmation run. \\
Test/fixture defect & The check failed because the newly-written test itself (its assertion or its synthetic fixture data) was wrong. \\
Production-logic defect & The check failed because the reverse-engineering/decoding logic under test was genuinely wrong. \\
Failed to propagate change to existing tests & The check failed because a correct change made elsewhere in the same session (a new method, a changed diagnostic message, a build-script fix) had not been carried through to a dependent mock, assertion, or script. \\
\bottomrule
\end{tabular}
\normalsize
\end{table}

\begin{table}[tb]
\footnotesize
\centering
\caption{Breakdown of Claude code generation event correctness, based on analysis of session transcripts.}\label{tab:selfcorrect}
\begin{tabular}{@{}
  >{\raggedright\arraybackslash}p{(\linewidth - 4\tabcolsep) * \real{0.6051}}
  >{\raggedleft\arraybackslash}p{(\linewidth - 4\tabcolsep) * \real{0.1824}}
  >{\raggedleft\arraybackslash}p{(\linewidth - 4\tabcolsep) * \real{0.2125}}
@{}}
\toprule
Category & Runs & \% \\
\midrule
Clean (no failure) & 90 & 85.7\% \\
Test/fixture defect & 8 & 7.6\% \\
Production-logic defect & 4 & 3.8\% \\
Failed to propagate change to existing tests & 3 & 2.9\% \\
\textbf{Total} & \textbf{105} & \textbf{100\%} \\
\bottomrule
\end{tabular}
\normalsize
\end{table}

Table \ref{tab:selfcorrect} shows that 15 out of 105 code generation events (14.3\%) introduced genuine coding errors, later corrected before any commit. Eight of these (7.6\% of all code generation events) were errors in the tests and four (3.8\%) were errors in the code itself. The remaining three (2.9\%) were failures to update tests after a (correct) code change.

Note that this analysis covers the 105 events for which a full session transcript is available and which fall within the paper's audited population window, due to imperfect record keeping\footnote{By default, session transcripts are deleted after 30 days and so some transcripts were lost before we discovered this.} not all code generation events could be recovered. Nevertheless, the data is enough to give a rough idea of the code generation error rate. In particular, that it is not small and so a good code test-suite is important for protecting against regressions.

To the best of our knowledge previous AI code quality analysis has all been based on commit history, and so is blind to bugs fixed before being committed. For the same reason, we are also not aware of a comparable human baseline since standard defect-density measurement (bug trackers, QA testing, release history) only observes code after it is committed. The human analogue would be errors caught and fixed within a developer's own write-test-fix iterations before committing, a rate existing measurement cannot see for the same reason.

\section{Reliability of Claude user interactions}\label{reliability-of-claude-user-interactions}

We evaluate the reliability of Claude's responses when interacting directly with the user. To do this we build a fact checking pipeline. For each response: (i) decompose it into component statements, (ii) decide whether each component contains a checkable fact or not, (iii) if it does, gather evidence to inform verification, and (iv) evaluate response component accuracy.

This decompose-classify-verify pipeline follows the same high-level architecture as existing automated fact-checking systems e.g.~ClaimBuster's four-stage pipeline (monitor, spot, match, check) {[}20{]}, SAFE's fact-decomposition-and-verification approach {[}21{]}. However, the details differ since checking factual claims about software is rare in the literature. As far as we know, we present the first fact-checking analysis of AI coding-session transcripts.

\subsection{Fact-checking pipeline}\label{fact-checking-pipeline}

\subsubsection{Categorise responses}\label{categorise-responses}

Parts (i) and (ii) of the pipeline are developed using a similar approach to \S\ref{sec:codebook}--\S\ref{sec:codebook-refinement}. We use an LLM (Codex GPT-5.6 Sol) to decompose into component parts each of the 678 responses to the 678 user instructions studied in \S \ref{sec:instructions}. This yields 9,967 response components. Of these, 6,818 components are a concluding paragraph, a bolded takeaway or a summary table i.e.~the elements of the response that a human would likely pay most attention to. We therefore focus on these 6,818 components in the analysis below.

We categorise the response components using an LLM (Codex GPT-5.6 Sol) instructed using: (i) the taxonomy summarised in Table\textasciitilde{}\ref{tab:response-category-definitions}, and (ii) the response component together with its immediate context (the preceding instruction and the AI's full response). The response taxonomy was developed using an iterative abductive approach: select categories iteratively with convergence after three rounds, then refine boundary cases by examining the differences between two independent LLMs (Claude Sonnet 5 and Codex GPT-5.6 Sol) categorisations. For the final codebook Cohen's \(\kappa=0.803\) (89.63\% raw agreement) between Claude and Codex, and \(\kappa=0.800/0.746\) between Claude/Codex and a human categorising a stratified sample of 5 responses from each category.

Prior claim taxonomies come mostly from journalism and social-media fact-checking, and classify by whether a claim is numeric, a quote, an entity/event property, a position statement {[}19{]}, and verification is by internet search. Our taxonomy also organises claims by the verification method each requires, but verification uses software artefacts. Checkable-fact and mechanism-explanation are settled using the commit the response was made against: by executing, searching, or reading code at that one point in time. Root-cause-diagnosis and design-fix-proposal instead require looking forward past that commit: whether a fix targeting the diagnosed cause went on to resolve the symptom, or whether the proposed change was actually implemented and still present at a later, fixed reference commit.

The full codebook, including definitions, key signals, examples, and boundary notes, is provided in our dataset package.

\begin{table*}[tb]
\footnotesize
\centering
\caption{Definitions of the response categories.}\label{tab:response-category-definitions}
\begin{tabular}{@{}
  >{\raggedright\arraybackslash}p{(\linewidth - 4\tabcolsep) * \real{0.1000}}
  >{\raggedright\arraybackslash}p{(\linewidth - 4\tabcolsep) * \real{0.6200}}
  >{\raggedright\arraybackslash}p{(\linewidth - 4\tabcolsep) * \real{0.2800}}
@{}}
\toprule
\begin{minipage}[b]{\linewidth}\raggedright
Category
\end{minipage} & \begin{minipage}[b]{\linewidth}\raggedright
Definition
\end{minipage} & \begin{minipage}[b]{\linewidth}\raggedright
Verdict categories
\end{minipage} \\
\midrule
Checkable-fact & A single fact about one artifact (a count, a presence/absence, a structural property) e.g.~``Full suite green: 1174 passed, 0 failed.'' & Matches / Contradicts / Unreproducible \\
Mechanism-explanation & Relates multiple facts or code locations into an explanation of how existing logic works (data flow, control flow, a multi-step process) e.g.~``The revert at extractor.py:4025-4028 fires when a field was tagged ENUM.'' & Confirmed / Wrong / Unreproducible \\
Root-cause-diagnosis & Diagnoses the cause of a stated or implied symptom/problem (something not working, unexpected, wrong, lost, or failing) e.g.~``The root cause: my scan uses \texttt{stop\_at\_ret=False} with a 256-insn window, which runs past the wrapper's end into neighboring functions, picking up other classes' cores.'' & Confirmed / Partial / Right-Conclusion-Wrong-Mechanism / Wrong / Unreproducible \\
Design-fix-proposal & Describes a proposed or implemented code change/fix and what happened to it e.g.~``The fix: attribute by control-flow block instead.'' & Adopted-Worked / Adopted-Revised / Adopted-Worked-But-Superseded / Not-Adopted / Failed / Mixed / Unreproducible \\
Other & A statement of intent, an interrogative, an opinion, a hedge, or pure process narration, e.g.~``The exact cause hasn't been pinned down yet.'' & Not scored \\
\bottomrule
\end{tabular}
\normalsize
\end{table*}

\begin{figure}[tb]
\centering
\includegraphics[width=0.48\linewidth,height=\textheight,keepaspectratio]{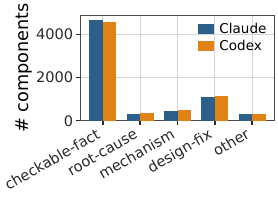}
\caption{Categories assigned to the response components.}\label{fig:category-distribution}
\end{figure}

Figure \ref{fig:category-distribution} shows the two independent LLMs distributions of categories. It can be seen that they are in good agreement. Checkable-fact dominates (\textasciitilde64-68\%), followed by Design-fix-proposal (16-17\%). Root-cause-diagnosis, Mechanism-explanation and Other each make up under 8\%.

\subsubsection{Evidence collection and scoring}\label{evidence-collection-and-scoring}

Table\textasciitilde{}\ref{tab:response-category-definitions} lists the possible verdicts for a response component in each category.
To support reasoning about the appropriate verdict for each response component, an evidence bundle is generated for each as follows.
Initially, for each response component a custom python script is used to gather an evidence bundle from the codebase itself (re-running code, static search, git-history tracing, binary parsing etc), then an LLM (Claude Sonnet 5) is used to score the response. The LLM has access to the codebase, session transcripts and project documents and is instructed to gather additional evidence as needed. Any such additional data is added to the evidence bundle and the response is re-scored. This is repeated until the evidence bundle converges (no new data is added): approximately 60\% converged immediately, 35\% with one extra round and 5\% with two extra rounds.

Using the converged evidence bundle, a stratified sample of 20 response components (five in each of the four categories) was scored by two independent LLMs (Claude Sonnet 5 and Codex GPT-5.6 Sol). A human researcher reviewed disagreements between the two LLMs, looking for patterns in the boundary cases. For commonly occurring patterns the scoring rubric was updated with clarifying text. The stratified sample was then expanded in stages (5\%, 40\%, 75\%, then the full 100\% of each category), with the same disagreement review and rubric-refinement cycle repeated at each stage, until every response component in every category had been scored by both LLMs.

The full rubric including examples and boundary notes is provided in our dataset package.

\subsection{Results}\label{results-3}

Since Checkable-fact dominates the response components, drawing samples uniformly at random tends to lead to relatively few samples of the other categories of response and so wide confidence intervals. We therefore use a stratified sampling approach, drawing extra samples from less frequent categories. Table \ref{tab:response-reliability} summarises the response accuracy scoring by two independent LLMs for the resulting sample of approximately 3,900 response components (out of the 6,818 total population)\footnote{Note that the Claude verdicts use the Claude categorisations and the Codex verdicts use the Codex categorisations. Responses categorised as Other are not scored. UNREPRODUCIBLE verdicts are not included in the reported figures.}. To the best of our knowledge these are the first reported error rates for AI responses during multi-turn code development sessions.

\begin{table}[tb]
\centering
\begin{minipage}{0.66\linewidth}
\centering
\scriptsize
\begin{tabular}{@{}
  >{\raggedright\arraybackslash}p{(\linewidth - 6\tabcolsep) * \real{0.20}}
  >{\raggedright\arraybackslash}p{(\linewidth - 6\tabcolsep) * \real{0.07}}
  >{\raggedleft\arraybackslash}p{(\linewidth - 6\tabcolsep) * \real{0.45}}
  >{\raggedleft\arraybackslash}p{(\linewidth - 6\tabcolsep) * \real{0.30}}
@{}}
\toprule
Category & Scorer & Accurate & 95\% CI \\
\midrule
Checkable-fact & Claude & 2321/2461 (94.3\%) & 93.3--95.2\% \\
 & Codex & 2305/2443 (94.4\%) & 93.4--95.2\% \\
Mechanism-explanation & Claude & 381/418 (91.1\%) & 88.0--93.5\% \\
 & Codex & 376/420 (89.5\%) & 86.2--92.1\% \\
Root-cause-diagnosis & Claude & 275/308 (89.3\%) & 85.3--92.3\% \\
 & Codex & 275/304 (90.5\%) & 86.6--93.3\% \\
Design-fix-proposal & Claude & 565/712 (79.4\%) & 76.2--82.2\% \\
 & Codex & 566/709 (79.8\%) & 76.7--82.6\% \\
\bottomrule
\end{tabular}
\end{minipage}
\hfill
\begin{minipage}{0.31\linewidth}
\centering
\includegraphics[width=\linewidth]{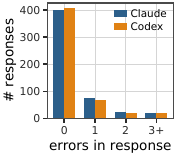}
\end{minipage}
\caption{Left: evaluation verdicts for each of the response categories, with Wilson 95\% confidence intervals. Right: number of responses containing 0, 1, 2, or 3+ inaccurate components (Checkable-fact, Root-cause-diagnosis and Mechanism-explanation categories only).}
\label{tab:response-reliability}
\label{fig:error-count-per-response}
\end{table}

Table \ref{tab:response-reliability} shows that there is good agreement between the two independent LLMs across all categories of response. Checkable-fact response components are the most accurate with approximately 6\% (1 in 18) containing errors. Mechanism-explanation and Root-cause-diagnosis are less accurate, with error rates of approximately 10\%. Design-fix-proposal has notably lower accuracy than the other categories, but this is also an inherently harder category to get right. Design-fix-proposal is a proposed code fix. Whether it is Adopted or Not-Adopted is decided by the user and involves a judgement call e.g.~on the cost/risk of the proposed change, on whether the benefit is sufficient to make the change worthwhile. The Not-Adopted cases are therefore not really response errors, yet the table's Accurate column counts them against accuracy since they are neither confirmed working nor Unreproducible. Excluding the 107 Not-Adopted cases (95 for Codex) from the denominator, and counting only Adopted-Revised, Mixed and Failed as errors (each was, at least in part, revised or abandoned because of a deficiency, or in the Failed case reverted outright), gives an accuracy of 565/605 (93.4\%) for Claude and 566/614 (92.2\%) for Codex.

A full response can contain multiple components, and this component error rate translates into 23.0\% of responses containing one or more inaccurate components for Claude and 21.6\% for Codex, i.e.~roughly 1 in 4-5 responses contain errors\footnote{These figures exclude Design-fix-proposal components, and responses where that is the only scoreable component. When included the error rate rises to 29.5\% for both Claude and Codex.}. Table \ref{fig:error-count-per-response} shows that the majority of responses have either zero or one error, but around 8\% of responses containing two or more errors.

A human researcher with access to all of the project artefacts manually checked 200 of the verdicts reporting deficiencies (Contradicts, Wrong etc) and a stratified sample of 200 clean verdicts (Matches, Confirmed etc), 49-51 per category. The accuracy between the AI and human verdicts is: Checkable-fact 98.2\%, Mechanism-explanation 98.0\%, Root-cause-diagnosis 96.6\%, Design-fix-proposal 93.3\%. The verdicts with AI-human disagreement were not modified as the aim here is to calibrate the AI results against human judgement. That said, if they were applied then the accuracy figures in Table \ref{tab:response-reliability} would change by: Checkable-fact 0\% (changes cancel out), Mechanism-explanation -0.48\%, Root-cause-diagnosis +0.25\%, Design-fix-proposal +0.60\%. These shifts are minor, well within the reported confidence intervals.

See Additional Material in our dataset package for examples of errors in each category.

\section{Limitations}\label{limitations}

\textbf{Scope: one dataset.}
Probably the main limitation of the work is that it is confined to a single dataset, albeit for a reasonably sized, real AI-written tool. On the other hand, full-project datasets, meaning commits plus session transcripts and AI-generated notes rather than just commits, are otherwise largely absent from the public record, so the dataset itself is one of the paper's primary contributions, which we hope will spur publication of other datasets in future.

\textbf{LLM-based categorisation and scoring.}
Much of the analysis relies on LLMs to decompose, categorise, and score instructions, commits, and responses. We use two independent LLMs (Claude and Codex, from two competing vendors) throughout, which consistently agree with each other and with human calibration samples (\(\kappa\) = 0.70--0.81), with no evidence of Claude favouring its own output. A sample of all (both defect/error and agreed-clean) verdicts was additionally checked by a human researcher with only a minor level of disagreement observed.

\section{Conclusions}\label{conclusions}

Our main contributions are: (i) the Extractor dataset, (ii) two reusable tools for tracing AI-authored Python code back to its provenance, (iii) three taxonomies, extending or filling a gap in an existing classification scheme; and the process, reliability, and communication findings these yield when applied to Extractor's development. We find that user coding agent CLI instructions differ in kind from IDE-chat instructions, with a greater focus on comprehension, planning and consultation, that AI code-generation errors are common (14.3\% of checked events introduced one), but the same AI process that made the mistakes also built the test suite that caught them before commit. Claude's responses are relatively reliable when reporting or verifying a fact (94.3\% accurate), when explaining an existing mechanism (91.1\%) or diagnosing a root cause (89.3\%) but less so when proposing a design fix (79.4\%). Overall roughly 1 in 4-5 responses contain at least one inaccuracy.

\section*{References}\label{references}
\addcontentsline{toc}{section}{References}

\protect\phantomsection\label{refs}

\section{Additional Material}\label{additional-material}

\subsection{Examples of response errors}\label{examples-of-response-errors}

\textbf{Checkable-fact.} \emph{``The top of the list is dominated by low-level disassembly plumbing (\texttt{find\_adrp\_ldr\_target}, \texttt{find\_adrp\_add\_target}, \texttt{\_is\_valid\_selector}, \texttt{decode\_ptr}), called millions of times during any parsing, not per-output-decision.''}

This claim is contradicted by the response's own work. Thirty-three seconds earlier, the AI had itself computed exact per-function call counts for a benchmark run across three test artifacts: the highest of the four named functions was called 296,681 times, the lowest 40,780. Two and a half minutes later it wrote these same counts into a report file, later committed to the repository. All four counts are well under 300,000, not ``millions''. A static count of call sites in the current codebase confirms the structural reason: each function scans a small, fixed-size instruction window and is invoked once per higher-level decision (once per enum case, once per dispatch table), not once per instruction.

\textbf{Mechanism-explanation.} \emph{``The golden diff is exactly the 9 fields I predicted, and none of them are \texttt{GEOPDComponentValue}/\allowbreak{}\texttt{GEOPDComponent}. So in the full \texttt{extract\_all} run those classes are unchanged (resolved via readFrom dispatch, not writeTo).''}

The fact half of this claim (which 9 fields changed) is correct, but the mechanism is stated backwards. Thirteen minutes later, in the same session, the AI debugs why four tests are still failing despite this claim, and concludes in its own words that ``the walk itself recovers the wide-oneof field \#91'', where ``the walk'' is the writeTo control-flow-graph walk it had just been editing, not readFrom dispatch. The commit implementing this fix is titled ``CFG walk is the sole writeTo decoder'' and its message credits the writeTo walk, not readFrom dispatch, with recovering exactly the class and field the claim names.

\textbf{Root-cause-diagnosis.} \emph{``That's because \texttt{/tmp/c72.txt} lives in my tool sandbox, not your shell's filesystem.''}

When the user reports that a file the AI had pointed them to could not be found, the AI diagnoses a filesystem boundary between its own tool environment and the user's shell. Twelve lines later, in the same session, the AI runs the identical command itself, from the very environment it says the file lives in, and gets the identical ``file not found'' error. It briefly offers a second explanation (that the file ``was wiped during the long wait'') before working around the problem with an inline heredoc instead. Neither explanation is checked, and neither is correct: earlier in the same session, every previous attempt to create the file had silently failed because of an unrelated tool outage, so the file had never existed anywhere. The AI disproves its own diagnosis without appearing to notice.

\textbf{Design-fix-proposal.} \emph{``Raising \texttt{\_SPARSE\_MAX\_INSNS} (currently 1000) would recover values for at least some of the larger enums, and it's the only constant that would.''}

Asked whether increasing a scan-length constant would help recover values from large enums that a linear scan currently truncates, the AI proposes raising it. The user asks for this to be tested before deciding, and the AI runs the constant at 1000, 4000 and 30,000 instructions against the full test corpus (752 enums, 14,900 values): every enum gains zero additional values at any of the three settings. The constant was not the cause. The actual bug, found and fixed the same day, was an unrelated jump-table decoding gap; a second, separate fix later that day corrects a false ``truncated'' warning on long methods that were already fully covered.

\end{document}